# Artificial Intelligence for Sustainable Urban Biodiversity: A Framework for Monitoring and Conservation

Yasmin Rahmati

Global Environment and Development

University of Copenhagen, Copenhagen, Denmark

Email:fmg425@alumni.ku.dk

**Abstract**— The rapid expansion of urban areas challenges biodiversity conservation, requiring innovative ecosystem management. This study explores the role of Artificial Intelligence (AI) in urban biodiversity conservation, its applications, and a framework for implementation. Key findings show that: (a) AI enhances species detection and monitoring, achieving over 90% accuracy in urban wildlife tracking and invasive species management; (b) integrating data from remote sensing, acoustic monitoring, and citizen science enables large-scale ecosystem analysis; and (c) AI decision tools improve conservation planning and resource allocation, increasing prediction accuracy by up to 18.5% compared to traditional methods. The research presents an AI-Driven Framework for Urban Biodiversity Management, highlighting AI's impact on monitoring, conservation strategies, and ecological outcomes. Implementation strategies include: (a) standardizing data collection and model validation, (b) ensuring equitable AI access across urban contexts, and (c) developing ethical guidelines for biodiversity monitoring. The study concludes that integrating AI in urban biodiversity conservation requires balancing innovation with ecological wisdom and addressing data quality, socioeconomic disparities, and ethical concerns.

1. ## Introduction

As biodiversity faces mounting pressures from global environmental change, urban areas present both challenges and opportunities for conservation (Weiskopf et al., 2024). The maintenance of urban biodiversity has become increasingly recognized as fundamental to both ecosystem functioning (Schwarz et al., 2017) and human welfare in cities (Taylor & Hochuli, 2015).

Balancing urban development with biodiversity conservation is, however, a complex task. The rapid pace of urbanization and associated human activities has exacerbated habitat loss, fragmentation, and the degradation of natural resources (Fraissinet et al., 2023). These pressures, driven by urban growth and densification, particularly over the past few decades, place significant strain on biodiversity and ecosystems (Horváth et al., 2019). Urban expansion frequently disrupts key ecological processes such as species movement, reproduction, and genetic exchange, jeopardizing the survival of many species (Donati et al., 2022).

While traditional conservation methods, including habitat restoration and protected areas, remain vital (Aronson et al., 2014), urban environments present unique challenges requiring adaptive strategies that integrate innovation and equity. The unique challenges of urban biodiversity stem from rapid urbanization, fragmented habitats, and anthropogenic disturbances, which demand more adaptive and innovative conservation strategies. (Seto et al., 2012)

Despite these challenges, urban areas can also serve as valuable sites for biodiversity conservation. Cities often occupy biologically rich regions (Ives et al., 2016) and host unique assemblages of flora and fauna, including rare and threatened species, within mosaics of semi-natural and artificial habitats (Jokimäki et al., 2018; Müller et al., 2013). By strategically planning and managing urban green spaces, cities can support biodiversity while enhancing ecosystem services for urban residents. (Spotswood et al., 2021)

Nature-based Solutions (NBS) have emerged as a promising framework for addressing these challenges, offering multi-benefit ecosystems that can simultaneously support biodiversity while providing environmental, social, and economic benefits (Cohen-Shacham et al., 2016), but the growing complexity of urban systems necessitates further innovation, particularly in leveraging cutting-edge technologies like artificial intelligence (AI) to enhance biodiversity outcomes.

The dynamic and fragmented nature of urban ecosystems requires innovative approaches for effective management. Local-scale planning decisions significantly influence the ecological outcomes of urban development, but these interventions must account for complex socio-ecological interactions (Brown, 2017).

Recent advances in AI and deep learning offer transformative potential in this regard. Modern technological solutions, especially AI and remote sensing, offer new capabilities for urban biodiversity monitoring and management at unprecedented scales. (Lausch et al., 2015). AI technologies enhance biodiversity assessment, habitat monitoring, and species preservation while analyzing the complex interactions between technological solutions and traditional conservation approaches (Ditria et al., 2022; Green et al., 2020). These advancements create synergies between innovation and established methods, bridging gaps in urban biodiversity management. For example, AI systems enhance species distribution mapping, corridor identification, and urban ecosystem assessment, enabling evidence-based conservation decisions (He et al., 2015).

The rapid advancement of AI capabilities and technologies has enabled its integration across numerous sectors including healthcare, manufacturing, education, public services, marketing and urban development. AI applications are transforming how organizations operate and make decisions (Dwivedi et al., 2021; Sadeghi & Niu, 2024). Recent work by Sadeghi and Niu (2024) demonstrated AI's capacity to address complex systemic challenges in education through a comprehensive framework for ethical implementation and human-AI collaboration. Building on such foundational approaches to AI deployment, urban biodiversity conservation presents unique opportunities where AI's transformative potential lies in its ability to bridge ecological challenges and human demands through real-time data processing, predictive analytics, and innovative modeling frameworks.

By using remote sensing data, for instance, AI-powered methods can identify patterns in urban biodiversity and socio-economic variables with remarkable precision (Grafius et al., 2019;

Prodanovic et al., 2024). Integrating AI into urban planning processes enables modeling workflows, facilitates effective stakeholder engagement, and provides advanced tools for visualizing project outcomes (Prodanovic et al., 2024). Moreover, AI applications make conservation more effective by addressing both environmental and socio-economic dimensions of urban biodiversity preservation (Kuller et al., 2019; Rega-Brodsky et al., 2022). These tools enable researchers and practitioners to monitor biodiversity dynamics, predict ecological changes, and design evidence-based strategies for urban planning and conservation (Zhou et al., 2023).

This study investigates the role of AI in urban biodiversity conservation and management. It explores how AI-driven technologies enhance species detection, habitat modeling, and urban ecosystem analysis, focusing on their applications in monitoring biodiversity, managing invasive species, and optimizing urban green spaces. Specifically, the research examines the integration of AI with traditional conservation methods to address urbanization challenges, such as habitat fragmentation and ecological degradation. Moreover, this study highlights the potential of AI to improve decision-making in urban planning through real-time monitoring, predictive analytics, and data-driven insights. By introducing a conceptual framework for AI applications in urban biodiversity management, the research provides practical guidance for policymakers, urban planners, and conservation practitioners. The study aims to bridge the gap between technological advancements and sustainable urban biodiversity strategies, offering insights into supporting resilient and biodiverse urban environments.

## 2. AI Applications in Urban Biodiversity Management

AI transforms urban biodiversity management by enabling species monitoring, habitat modeling, and ecosystem planning. Key applications include (a) Species Detection and Monitoring, (b) Urban Ecosystem Analysis and Planning, (c) Acoustic and Environmental Monitoring, (d) Habitat and Species Distribution Modeling, and (e) Conservation Planning and Management. Table 1 summarizes recent studies (2020–2024) showcasing AI applications across these themes.

### 2.1 Species Detection and Monitoring

The application of AI in urban biodiversity research has transformed species detection and monitoring, enabling precise, scalable, and cost-effective ecosystem management. Recent advancements highlight deep learning's effectiveness in improving the accuracy, scalability, and efficiency of monitoring across various taxa and environments.

In urban wildlife monitoring, deep learning models have achieved notable success in detecting and managing animal populations. For instance, the Swin-Mask R-CNN with SAHI model developed for feral pigeon detection in Hong Kong significantly improved monitoring precision. By utilizing Swin Transformers for feature extraction and the SAHI tool for small object detection, the model achieved a 10% precision improvement, providing a scalable solution for automated wildlife monitoring (Guo et al., 2024).

In vegetation monitoring, AI-based approaches have advanced urban flora mapping. A multi-task convolutional neural network (CNN) for tree species mapping in Rio de Janeiro achieved high classification accuracy, with F1-scores of 79.3% for nine species and 87.6% for five dominant species. This method addressed challenges such as overlapping canopies and tree trait variability, providing valuable tools for urban forest management and green infrastructure planning (Martins et al., 2021). In invasive species management, combining UAV imagery with the DeepLabv3+ model successfully monitored the invasive aquatic plant Pistia stratiotes in urban water bodies, achieving 90.24% accuracy. This enabled early detection and intervention in dynamic aquatic ecosystems, despite challenges like vegetation occlusion and environmental variability (Hao et al., 2024).

## 2.2 Urban Ecosystem Analysis and Planning

The integration of AI and advanced technologies has transformed urban ecosystem analysis and planning, providing innovative solutions to balance urban development with biodiversity conservation amid challenges like habitat fragmentation, pollution, and climate change.

GIS-based tools like the Habitat Network Analysis Tool (HNAT) automate habitat network analyses to predict species distribution and assess functionality. Tested in Gothenburg, Sweden, HNAT integrated habitat quality and connectivity data, identifying areas for restoration to improve amphibian habitats while incorporating factors like road networks and traffic for effective biodiversity planning (Kindvall et al., 2024).

AI-driven remote sensing technologies, including NDVI, satellite imagery, and LiDAR, have enhanced large-scale urban vegetation and biodiversity mapping. The ECO-LENS project used these tools with machine learning to analyze NDVI trends in 65 cities, linking state-level policies to vegetation coverage and supporting collaborative conservation efforts, despite challenges like data labeling and NDVI's species-specific limitations (Montas, 2024).

AI also optimizes green space management through systems like the "Green Space Optimizer," which combines real-time environmental data and predictive analytics to support biodiversity, water conservation, and community engagement. Its ability to mitigate urban heat island effects and improve carbon footprints makes it a global benchmark for urban park management (Patil et al., 2024).

Citizen science and clustering further enhance biodiversity planning. In Athens, Greece, a k-means clustering approach combined remote sensing data with citizen science observations from GBIF, classifying urban habitats at the zipcode level to identify biodiversity hotspots and emphasize green infrastructure in planning. This replicable framework highlights public engagement's value in conservation (Ziliaskopoulos & Laspidou, 2024).

These advancements demonstrate AI's transformative potential in urban ecosystem analysis and planning by integrating GIS, remote sensing, machine learning, and citizen science to create sustainable, resilient urban environments.

## 2.3 Environmental Acoustics and Soundscape Analysis

Environmental acoustics and soundscape analysis, using tools like eco-acoustic indices, passive acoustic monitoring (PAM), and machine learning, offer non-invasive and scalable methods to monitor species diversity, habitat quality, and human impact on urban ecosystems.

In their study, Latifi et al. (2023) showed the effectiveness of eco-acoustic indices—Acoustic Complexity Index (ACI), Bioacoustics Index (BI), and Normalized Difference Soundscape Index (NDSI)—in assessing bird biodiversity in Isfahan, Iran. Specifically, parks with less noise and better vegetation, such as Soffeh Park, showed higher bird species richness and activity. Through their analysis, machine learning models like Support Vector Machine (SVM) and Random Forest (RF) predicted biodiversity indices with high accuracy, with SVM achieving $R^2$ values of 0.93 for songbird richness and 0.92 for evenness. Taken together, these results highlight the value of acoustic monitoring and thoughtful park design in enhancing urban habitat quality.

Building on these acoustic monitoring advances, Zhang et al. (2023) used deep learning to classify acoustic scenes in Guangzhou's urban forests, analyzing seven sound categories, including human and animal sounds. In their implementation, the DenseNet_BC_34 model achieved 93.81% accuracy, with mel spectrograms effectively capturing temporal and spectral features. However, challenges like misclassification in mixed soundscapes due to overlapping features emphasized the need for diverse training datasets. Ultimately, this study showcases deep learning's role in advancing passive acoustic monitoring for urban biodiversity research.

## 2.4 Species Distribution and Habitat Modeling

AI-driven species distribution and habitat modeling has become a key tool in urban biodiversity research, providing precise methods to assess species-environment interactions using Geographic Information Systems (GIS), Earth Observation (EO) data, and machine learning. For instance, Zheng et al. (2024) used GIS and machine learning to assess urbanization's impact on bobcat habitats in San Jose, California. Their innovative Habitat Suitability Model (HSM) integrated vegetation cover, water distribution, road traffic, and intersection density to identify critical conservation areas. Through detailed analysis, the study found that while vegetation and water bodies improved habitat suitability, proximity to busy roads diminished it, ultimately emphasizing the need for wildlife crossings to enhance connectivity and reduce roadkill. This comprehensive framework informs local wildlife strategies and broader urban ecological challenges.

Similarly focusing on urban habitats, Wellmann et al. (2020) used high-resolution EO data and machine learning to model habitats for 44 bird species in Leipzig, Germany, achieving accuracies of 59–90%. Notably, continuous vegetation indicators, such as density and texture metrics, outperformed traditional indices like NDVI in predicting habitat suitability. By combining individual Species Distribution Models (SDMs), they successfully mapped urban bird species richness, thereby offering guidance for optimizing urban green spaces to support biodiversity.

Expanding on these findings, Zhai et al. (2024) analyzed spatiotemporal bird species richness patterns in Beijing, linking urbanization, seasonal changes, and environmental factors using algorithms like Random Forest and Extreme Gradient Boosting. Their research revealed that key predictors included water bodies, green space area, relative humidity, and nighttime light pollution. Most significantly, suburban areas with blue-green infrastructure saw increased diversity, while core urban areas experienced declines, further underscoring the importance of integrating blue-green networks into urban planning.

Finally, complementing these urban studies, Zheran et al. (2020) demonstrated EO-based SDMs' scalability in urban avifauna assessment. Through their methodological approach linking spectral vegetation traits and vegetation heterogeneity to bird behaviors, they developed a cost-effective framework for modeling species richness and supporting urban biodiversity management.

### 2.5 Conservation Planning and Management

AI applications in conservation planning have transformed biodiversity protection, invasive species management, and urban forest enhancement by leveraging machine learning, remote sensing, and advanced analytics.

Silvestro et al. (2022) introduced CAPTAIN, an AI-based framework using reinforcement learning to dynamically prioritize conservation areas. CAPTAIN outperformed traditional tools like Marxan, achieving up to 18.5% lower species loss under budget constraints. By integrating real-time biodiversity data, anthropogenic disturbance metrics, and climate change projections, CAPTAIN achieved conservation targets in 68% of simulations for Madagascar's endemic trees, compared to Marxan's 2%, demonstrating its efficiency in balancing diverse conservation goals.

For invasive species management, Dutta et al. (2020) applied ResNet-18, a deep learning model, to detect Elaeagnus umbellata in urban parks. With over 96% accuracy across datasets from eight parks, the model's scalability enables dynamic monitoring and resource-efficient mitigation in metropolitan areas, offering critical guidance for targeted invasive species control.

In urban forestry, Louis et al. (2022) used machine learning techniques like PCA and boosting regression to analyze tree biodiversity along Hong Kong's San Tin Highway. Slopes supported higher biodiversity than verges dominated by Corymbia citriodora, with slope height and area as key predictors of biodiversity. This approach provided strategies to balance biodiversity enhancement with safety concerns, such as wildlife-vehicle collisions and risks from tall trees.

These studies illustrate how AI-driven tools optimize resource use, improve biodiversity outcomes, and offer adaptive strategies to address urbanization and ecological challenges.

### 2.6 Urban Environmental Change Analysis

AI applications in urban environmental change analysis integrate machine learning, remote sensing, and IoT technologies to monitor, model, and manage urbanization's impact on biodiversity, offering actionable insights for conservation and sustainable urban development. For example, Wang et al. (2022) demonstrated the potential of AIoT systems for real-time water quality monitoring in urban water bodies. By combining IoT sensors with machine learning models like General Regression Neural Network (GRNN) and Multivariate Polynomial Regression

(MPR), the study achieved highly accurate predictions of water quality parameters, with errors below 0.2 mg/L. This cost-effective approach facilitates efficient monitoring and pollutant management, highlighting AI's role in safeguarding aquatic ecosystems.

Furthermore, Eyster et al. (2024) analyzed a 26% decline in bird abundance in Metro Vancouver from 1997 to 2020 using deep learning models to classify landcover changes. Notably, habitat measured at species-specific scales better explained temporal population changes, though broader drivers like climate change and arthropod declines also contributed. This comprehensive study underscores the importance of multiscale ecological assessments and habitat restoration to mitigate biodiversity loss.

Similarly, in the context of urban forestry, Elmes et al. (2018) addressed urban tree mortality in Worcester, MA, using Conditional Inference Trees (CIT) to identify socioeconomic and biophysical predictors in a replanting program. Their analysis revealed that factors such as renter occupancy, impervious surfaces, and property characteristics influenced tree survival, ultimately emphasizing the integration of socioeconomic variables into urban forestry strategies to enhance survivorship and ecosystem services.

**Table 1:** Analysis of AI applications in urban biodiversity studies (2020-2024)

| Theme | Study Authors (Year) | AI Methods/ Tools Used | Data Types | Key Findings | Accuracy/ Performance | Context/ Location |
|---|---|---|---|---|---|---|
| **Species Detection and Monitoring** | Guo et al. (2024) | Swin-Mask R-CNN with SAHI | High-resolution images | Improved pigeon detection and counting in complex urban environments | 74% mAP; 10% increase in AP50s for small targets | Hong Kong urban areas |
|  | Martins et al. (2021) | Multi-task CNN | Aerial photographs (RGB) | Successful mapping of urban tree species | F1-scores: 79.3% (9 species), 87.6% (5 species) | Rio de Janeiro urban forests |
|  | Hao et al. (2024) | DeepLabv3+ with ResNet-34 | UAV imagery | Effective monitoring of invasive aquatic plants | Average accuracy 90.24% | Urban water bodies |
| **Environmental Acoustics and Soundscape Analysis** | Zhang et al. (2023) | DenseNet_BC_34 | Audio recordings | Classification of urban forest acoustic scenes | 93.81% validation accuracy | Urban forests |
|  | Latifi et al. (2023) | SVM, RF, GLMNET | Acoustic indices | Bird biodiversity assessment in urban parks | R² values: 0.93 (songbird), 0.92 (evenness) | Urban parks |
| **Urban Ecosystem Analysis and Planning** | Wellmann et al. (2020) | Random Forest | Earth Observation data | Urban bird habitat modeling | Accuracies 59-90% | Urban bird habitats |
|  | Zhai et al. (2024) | DT, RF, XGBoost | Geographic and environmental data | Patterns of bird species distribution and richness | RF model achieved R² = 0.93, RMSE = 12.46 | Urban built-up areas in Beijing |
|  | Montas, Enrique B. (2024) | SAM, MobileNetV2, PointNet, KMeans | NDVI, satellite imagery, LiDAR data | Remote sensing and machine learning for biodiversity assessment | LiDAR achieved 83% accuracy | U.S., Colombia, Mexico biodiversity hotspots |

|  | Ziliaskopoulos, Konstantinos et al. (2024) | K-means clustering | Remote sensing data, GBIF species data | Framework identifies urban habitat types and emphasizes green infrastructure | Not applicable (clustering results validated) | Athens, Greece |
| --- | --- | --- | --- | --- | --- | --- |
| **Species Distribution and Habitat Modeling** | Zheng et al. (2024) | GIS Tools, ML, Weighted Overlay | Geospatial data | Identified major impacts of urbanization on bobcat habitats | Performance not quantified | San Jose, California |
|  | Wellmann et al. (2020) | Machine Learning | High-resolution EO data | Urban bird species distribution modeling | Accuracies 59-90% | Leipzig, Germany |
| **Conservation Planning and Management** | Silvestro et al. (2022) | Reinforcement Learning (CAPTAIN) | Simulated + biodiversity data | Dynamic conservation planning, reduced species loss | Outperformed Marxan; Protected 26% more species | Non-urban (Madagascar biodiversity) |
|  | Dutta et al. (2020) | ResNet-18 | High-resolution aerial imagery | Robust detection of invasive plant species | 96.2%–97.6% accuracy | Urban parks in Charlotte, NC |
|  | Louis, Lee et al. (2022) | PCA, Linear Regression, Model Boosting | Biodiversity indices, geophysical data | Slope geophysical variables enhance biodiversity modeling | Adjusted $R^2$ increased by 0.23; RMSE reduced by 0.40 | Hong Kong roadside habitats |
| **Urban Environmental Change Analysis** | Wang et al. (2022) | GRNN, MPR, IoT-based sensors | Historical + IoT water quality data | AIoT-based water quality monitoring system | GRNN: MSE 0.91–1.11; MPR: $R^2$ 0.78–0.89 | Lam Tsuen River, Hong Kong |
|  | Eyster et al. (2024) | Bayesian models, boosted regression | Longitudinal bird surveys | Space-for-time exaggeration in bird-habitat relationships | Scale-optimized models improved prediction | Metro Vancouver, Canada |
|  | Elmes et al. (2024) | CIT, Logistic Regression | Tree mortality + sociodemographic data | Identified predictors of juvenile tree mortality in urban settings | Not applicable | Worcester, MA, USA |

## 3. AI-Driven Framework for Biodiversity Conservation and Management

The AI-Driven Framework for Biodiversity Conservation and Management (Fig. 1) provides a systematic approach to integrating AI technologies into conservation efforts through five hierarchical layers, transforming data into actionable outcomes. At the top, the AI Technologies & Tools Layer acts as the technological foundation guiding the framework. In designing this hierarchical framework, we drew inspiration from recent advances in AI implementation frameworks across different domains. Notably, Sadeghi and Niu (2024) established key principles for responsible AI deployment in complex social systems, emphasizing data standardization, ethical considerations, and the critical balance between automation and human oversight. These foundational concepts informed our approach to integrating AI in biodiversity conservation, particularly in ensuring equitable access to technology and maintaining transparency in decision-making processes.

The Data Management Layer incorporates five key data streams—Remote Sensing, Environmental Sensors, Acoustic Monitoring, Citizen Science, and Historical Records—to ensure comprehensive biodiversity indicators across scales. Building on this, the Analysis & Modeling Layer applies advanced tools such as Species Detection, Habitat Suitability Modeling, Ecosystem Service Assessment, Risk Analysis, and Predictive Modeling to generate insights into ecosystem dynamics and species populations.

Insights from analysis are operationalized in the Implementation & Monitoring Layer through Real-time Monitoring, Adaptive Management, Conservation Planning, and Impact Assessment, bridging analysis and action for data-driven, practical conservation. Finally, the Outcomes Layer focuses on four objectives: Enhanced Biodiversity, Ecosystem Resilience, Sustainable Urban Development, and Evidence-based Policy Making.

The framework functions as an integrated system, with sequential information flow and feedback loops enabling continuous refinement of strategies. This structured approach ensures AI is effectively leveraged to address biodiversity challenges while providing clear pathways for implementation and evaluation. It serves as both a practical guide for conservation managers and a conceptual model for researchers, emphasizing AI's transformative role in achieving conservation goals and sustainable ecosystem management.

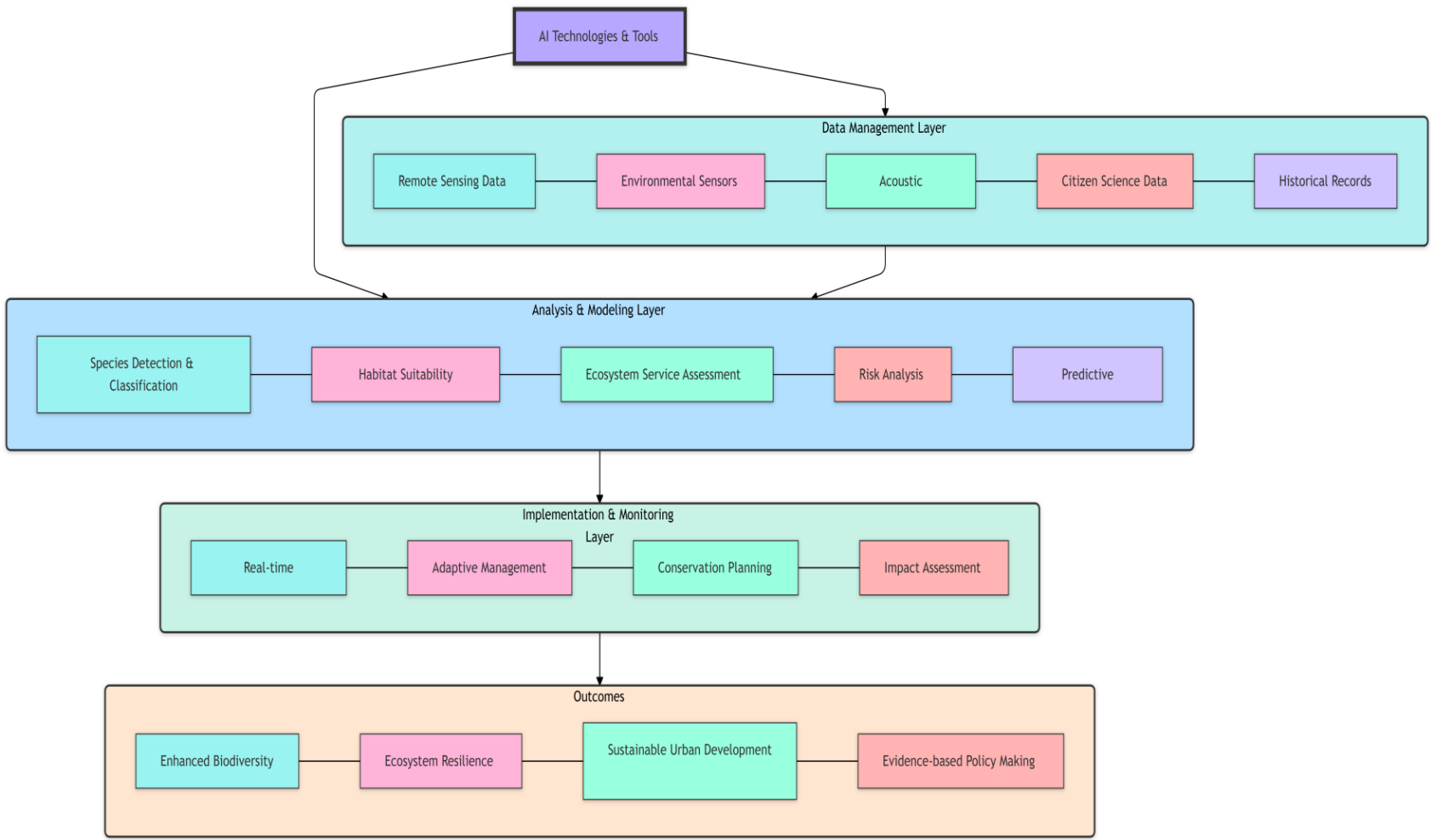

**Figure 1:** Hierarchical Framework for AI-Enabled Biodiversity Conservation and Management

## 4. Opportunities and Challenges in AI-Driven Urban Biodiversity Management

The integration of AI technologies in urban biodiversity conservation presents both significant opportunities and complex challenges that must be carefully considered. The rapid advancement of AI tools, particularly in species detection, habitat modeling, and environmental monitoring, has created unprecedented possibilities for understanding and protecting urban ecosystems. However, these opportunities are accompanied by substantial technical, ethical, and socioeconomic challenges that require thoughtful consideration and strategic solutions.

Remote sensing and AI-powered monitoring systems have demonstrated remarkable potential for scalable, cost-effective biodiversity assessment. As evidenced by studies like Guo et al. (2024), AI models can achieve detection accuracies exceeding 74% for urban wildlife, while maintaining real-time monitoring capabilities that were previously impossible with traditional methods. The integration of multiple data sources, including acoustic monitoring, satellite imagery, and citizen science observations, has enabled comprehensive ecosystem analysis at unprecedented scales. This scalability is particularly valuable in urban environments where rapid environmental changes demand continuous monitoring and adaptive management strategies.

The potential for integration with nature-based solutions (NBS) represents another significant opportunity. AI tools can optimize the design and implementation of green infrastructure, as demonstrated by the "Green Space Optimizer" system, which enhances urban park management for biodiversity while addressing climate resilience. These technologies enable precise targeting of conservation efforts, allowing cities to maximize biodiversity benefits within limited urban spaces while simultaneously addressing challenges like the urban heat island effect and carbon footprint reduction.

However, several critical challenges must be addressed. Data quality and availability remain persistent concerns, particularly in developing urban areas. The effectiveness of AI models depends heavily on comprehensive, high-quality training data, which is often lacking for many urban species and ecosystems. Privacy concerns also emerge when monitoring systems collect data in populated areas, requiring careful consideration of data governance and protection frameworks.

Ethical considerations and potential biases in AI models present another significant challenge. As highlighted by studies in species distribution modeling, AI systems may inherit biases from training data, potentially overlooking rare species or underrepresenting certain urban habitats. These biases can lead to skewed conservation priorities and ineffective management strategies if not properly addressed. The challenge of ensuring fair and equitable representation of different urban ecosystems in AI-driven decision-making processes requires ongoing attention and mitigation strategies.

Socioeconomic disparities in technology adoption and implementation pose additional challenges. While advanced AI tools offer powerful capabilities for biodiversity management, their deployment often requires significant technical expertise and financial resources. This disparity can lead to uneven implementation across different urban areas, potentially exacerbating existing inequalities in urban biodiversity conservation. Studies like Louis et al. (2022) highlight how socioeconomic factors significantly influence the success of urban greening initiatives, emphasizing the need for inclusive approaches to technology deployment.

Looking forward, addressing these challenges requires a multi-faceted approach. Investment in comprehensive data collection systems, development of standardized protocols for AI model validation, and establishment of clear ethical guidelines for urban biodiversity monitoring are essential steps. Additionally, promoting partnerships between research institutions, local governments, and community organizations can help bridge the technology gap and ensure more equitable access to AI-driven conservation tools. Success in urban biodiversity conservation ultimately depends on balancing the transformative potential of AI technologies with careful consideration of their limitations and societal implications.

## 5. Policy Implications and Recommendations

The integration of AI technologies in urban biodiversity conservation requires thoughtful policy frameworks to ensure effective and equitable outcomes (Guilherme et al., 2024); Rega-Brodsky et al., 2022). Based on the evidence presented, several key recommendations emerge for policymakers, urban planners, and conservation managers.

Investing in AI-driven biodiversity tools is a critical policy priority. Successful implementations like CAPTAIN (Silvestro et al., 2022) and AIoT water quality systems (Wang et al., 2022) demonstrate AI's ability to enhance conservation outcomes while optimizing resources. Policymakers should establish funding mechanisms for developing, piloting, and scaling AI tools, as evidenced by the Habitat Network Analysis Tool (HNAT) in Gothenburg, which yielded significant benefits for biodiversity planning (Kindvall et al., 2024).

The implementation of AI in biodiversity conservation raises critical questions about stakeholder engagement and system transparency. Research on AI implementation highlights that successful adoption relies on clear communication and inclusive stakeholder involvement (Sadeghi, 2024). In urban biodiversity contexts, this principle underscores the importance of emphasizing transparency about AI systems, particularly when engaging community groups, local organizations, and citizen scientists. Transparent communication improves trust and mitigates resistance to technological change, and ensures equitable participation in conservation efforts. For instance, studies in Athens demonstrated that meaningful community engagement in AI-driven biodiversity monitoring significantly improved the identification of biodiversity hotspots (Ziliaskopoulos & Laspidou, 2024).

Ensuring equitable access to AI technologies is crucial for inclusive biodiversity management. Policies must address the digital divide by building local capacity, providing technical training, and supporting technology adoption in underserved areas. In addressing the complexities of resource allocation for biodiversity projects, methodologies like the Sustainable, Robust, Resilient, and Risk-averse Budget Allocation for Projects (S3RBAP) framework (Lotfi et al., 2024) provide valuable insights. By incorporating risk mitigation, resiliency, and sustainability, such approaches can inform budget allocation strategies that balance ecological goals with stakeholder needs and resource constraints in urban biodiversity conservation. Studies in Worcester, MA, highlight how socioeconomic factors influence tree survival and conservation outcomes, underscoring the need for equity-focused technology deployment (Elmes et al., 2018).

Balancing conservation with urban development requires integrating biodiversity data into planning processes. Research on bobcat habitats in San Jose (Zheng et al., 2024) and bird distribution in Leipzig (Wellmann et al., 2020) emphasizes the importance of using AI-driven tools to guide urban planning while prioritizing ecological sustainability.

These strategies underscore the importance of AI in shaping policies that protect biodiversity, support urban sustainability, and promote equitable access to conservation technologies. The findings highlight the importance of ethical AI principles, as outlined by Sadeghi and Niu (2024), in promoting equitable and transparent conservation practices. This includes developing policies that:

- Mandate biodiversity impact assessments using AI-powered monitoring systems, following successful models like the DeepLabv3+ implementation (Hao et al., 2024)
- Establish minimum requirements for green space preservation based on AI-derived habitat modeling (Wellmann et al., 2020)
- Create incentives for developers to incorporate biodiversity-friendly design elements, informed by studies like Louis et al. (2022) on urban forestry
- Set clear metrics for measuring and monitoring conservation outcomes using advanced monitoring systems like those developed by Guo et al. (2024)

To implement these recommendations effectively, several specific policy actions are proposed, drawing from successful implementations documented in recent research:

1. Establish a dedicated urban biodiversity technology fund to support AI tool development and deployment, following models that have shown success in species monitoring (Martins et al., 2021)
2. Create regulatory frameworks for standardized data collection and sharing protocols, building on experiences from successful acoustic monitoring programs (Zhang et al., 2023)
3. Develop certification programs for AI-powered biodiversity assessment tools, incorporating lessons from successful implementations like the ECO-LENS project (Montas, 2024)
4. Implement mandatory biodiversity monitoring requirements for urban development projects, informed by successful models like those studied by Eyster et al. (2024)

5. Create incentive programs for private sector investment in biodiversity conservation technology, following successful examples of public-private partnerships (Latifi et al., 2023)

The success of these policy recommendations relies on sustained commitment and coordinated action across governance levels, as shown by successful urban biodiversity initiatives (Cohen-Shacham, 2016; Spotswood et al., 2021). Regular evaluation and adaptation based on monitoring outcomes and emerging technologies will be key to long-term effectiveness. Policies must also emphasize transparent decision-making and meaningful community engagement in conservation planning, as advocated by Brown (2017) and Rega-Brodsky et al. (2022).

As urban areas expand under increasing environmental pressures, these recommendations provide a framework for leveraging AI to enhance biodiversity conservation while promoting sustainable urban development. Integrating nature-based solutions and sustainable urban planning with AI frameworks highlights the value of combining technological innovation with holistic approaches to achieve ecological, social, and economic sustainability (Kabisch et al., 2016; (Tzoulas et al., 2007). The future of urban biodiversity conservation depends on combining AI tools with established practices to address urban-specific challenges. Effective implementation requires careful consideration of local contexts, resources, and policy frameworks to ensure equitable and impactful outcomes.

## 6. Conclusion and Future Research Directions

This study highlights the transformative potential of AI in urban biodiversity conservation, offering tools to monitor species, assess habitats, and enhance ecosystem resilience while supporting sustainable development. By replacing reactive methods with data-driven insights and real-time monitoring, AI enables proactive biodiversity management. AI analyzes vast datasets to uncover previously inaccessible ecosystem patterns. Through predictive modeling, real-time assessments, and automated monitoring, it equips policymakers and practitioners with tools for informed decisions, positioning AI as a catalyst for ecological sustainability. By integrating ecological goals with socio-economic considerations, AI reshapes urban planning, enhancing interdisciplinary collaboration and bridging development with conservation while empowering stakeholders to share responsibility for biodiversity.

However, AI implementation must address challenges like equitable access, ethical concerns, and over-reliance on technology. Partnerships, regulations, and community engagement are essential to aligning AI with diverse ecosystem needs.

Amid urbanization, AI redefines the relationship between cities and nature, advancing harmony between human activity and thriving ecosystems through balanced biodiversity strategies.

The proposed framework for integrating AI into urban biodiversity conservation provides a foundation for addressing urbanization and ecological challenges. Future research should test its applicability across diverse ecological, social, and economic settings to identify improvements and adaptations, refining it with emerging AI technologies, advanced analytics, socio-economic variables, real-time monitoring, and citizen science applications. Moreover, researchers can refine the framework by incorporating emerging AI technologies, advanced analytics, socio-economic variables, real-time monitoring, and citizen science applications. Collaboration with practitioners and stakeholders will ensure it remains practical and achieves equitable outcomes.

Furthermore, bibliometric analyses (Niu et al., 2024) can advance this research by mapping AI-biodiversity trends, identifying knowledge gaps, and evolving the framework into a dynamic tool for urban biodiversity management and sustainable conservation.

# References

Aronson, M. F., La Sorte, F. A., Nilon, C. H., Katti, M., Goddard, M. A., Lepczyk, C. A.,…Clarkson, B. (2014). A global analysis of the impacts of urbanization on bird and plant diversity reveals key anthropogenic drivers. *Proceedings of the royal society B: biological sciences*, *281*(1780), 20133330.

Brown, I. T. (2017). Managing cities as urban ecosystems: Fundamentals and a framework for Los Angeles, California. *Cities and the Environment (CATE)*, *10*(2), 4.

Cohen-Shacham, E., Walters, G., Janzen, C., & Maginnis, S. (2016). Nature-based solutions to address global societal challenges. *IUCN: Gland, Switzerland*, *97*, 2016-2036.

Donati, G. F., Bolliger, J., Psomas, A., Maurer, M., & Bach, P. M. (2022). Reconciling cities with nature: Identifying local Blue-Green Infrastructure interventions for regional biodiversity enhancement. *Journal of Environmental Management*, *316*, 115254.

Ditria, E. M., Buelow, C. A., Gonzalez-Rivero, M., & Connolly, R. M. (2022). Artificial intelligence and automated monitoring for assisting conservation of marine ecosystems: A perspective. *Frontiers in Marine Science*, *9*, 918104.

Dutta, D., Chen, G., Chen, C., Gagné, S. A., Li, C., Rogers, C., & Matthews, C. (2020). Detecting plant invasion in urban parks with aerial image time series and residual neural network. Remote Sensing, 12(21), 3493. https://doi.org/10.3390/rs12213493

Dwivedi, Y. K., Hughes, L., Ismagilova, E., Aarts, G., Coombs, C., Crick, T., ... & Williams, M. D. (2021). Artificial Intelligence (AI): Multidisciplinary perspectives on emerging challenges, opportunities, and agenda for research, practice and policy. *International journal of information management*, *57*, 101994.

Elmes, A., Rogan, J., Roman, L. A., Williams, C. A., Ratick, S. J., Nowak, D. J., & Martin, D. G. (2018). Predictors of mortality for juvenile trees in a residential urban-to-rural cohort in Worcester, MA. Urban Forestry & Urban Greening, 30, 138–151. https://doi.org/10.1016/j.ufug.2018.01.024

Eyster, H. N., Chan, K. M. A., Fletcher, M. E., & Beckage, B. (2024). Space-for-time substitutions exaggerate urban bird–habitat ecological relationships. Journal Name, Volume(Issue), page range. https://doi.org/10.1111/1365-2656.14194

Fraissinet, M., Ancillotto, L., Migliozzi, A., Capasso, S., Bosso, L., Chamberlain, D. E., & Russo, D. (2023). Responses of avian assemblages to spatiotemporal landscape dynamics in urban ecosystems. *Landscape Ecology*, *38*(1), 293-305.

Grafius, D. R., Corstanje, R., Warren, P. H., Evans, K. L., Norton, B. A., Siriwardena, G. M., ... & Harris, J. A. (2019). Using GIS-linked Bayesian Belief Networks as a tool for modelling urban biodiversity. *Landscape and Urban Planning*, *189*, 382-395.

Green, S. E., Rees, J. P., Stephens, P. A., Hill, R. A., & Giordano, A. J. (2020). Innovations in camera trapping technology and approaches: The integration of citizen science and artificial intelligence. *Animals*, *10*(1), 132.

Guo, Z., He, Z., Lyu, L., Mao, A., Huang, E., & Liu, K. (2024). Automatic Detection of Feral Pigeons in Urban Environments Using Deep Learning. Animals, 14(1), 159. https://doi.org/10.3390/ani14010159

Guilherme, F., Vicente, J. R., Carretero, M. A., & Farinha-Marques, P. (2024). Mapping multigroup responses to land cover legacy for urban biodiversity conservation. *Biological Conservation*, *291*, 110508.

Hao, Z., Lin, L., Post, C. J., & Mikhailova, E. A. (2024). Monitoring the spatial–temporal distribution of invasive plant in urban water using deep learning and remote sensing technology. Ecological Indicators, 162, 112061. https://doi.org/10.1016/j.ecolind.2024.112061classification

He, K. S., Bradley, B. A., Cord, A. F., Rocchini, D., Tuanmu, M. N., Schmidtlein, S., ... & Pettorelli, N. (2015). Will remote sensing shape the next generation of species distribution models? Remote Sensing in Ecology and Conservation, 1(1), 4–18.

Horváth, Z., Ptacnik, R., Vad, C. F., & Chase, J. M. (2019). Habitat loss over six decades accelerates regional and local biodiversity loss via changing landscape connectance. *Ecology letters*, *22*(6), 1019-1027.

Ives, C. D., Lentini, P. E., Threlfall, C. G., Ikin, K., Shanahan, D. F., Garrard, G. E.,…Rayner, L. (2016). Cities are hotspots for threatened species. *Global Ecology and biogeography*, *25*(1), 117-126.

Jokimäki, J., Suhonen, J., & Kaisanlahti-Jokimäki, M.-L. (2018). Urban core areas are important for species conservation: A European-level analysis of breeding bird species. *Landscape and urban planning*, *178*, 73-81.

Kabisch, N., Stadler, J., Korn, H., & Bonn, A. (2016). Nature-based solutions to climate change mitigation and adaptation in urban areas: Perspectives on indicators, knowledge gaps, and opportunities for action. Ecological Indicators, 69, 581–596.

Kellenberger, B., Marcos, D., & Tuia, D. (2018). Detecting animals in UAV images: Best practices to address a substantially imbalanced dataset with deep learning. *Remote Sensing of Environment, 216*, 139–153.

Kindvall, O., Berghauser Pont, M., Stavroulaki, I., Lanemo, E., Wigren, L., & Levan, M. (2024). Predicting habitat functionality using habitat network models in urban planning. Environment and Planning B: Urban Analytics and City Science, 0(0). https://doi.org/10.1177/23998083241299165

Kuller, M., Bach, P. M., Roberts, S., Browne, D., & Deletic, A. (2019). A planning-support tool for spatial suitability assessment of green urban stormwater infrastructure. *Science of the total environment*, *686*, 856-868.

Latifi, M., Fakheran, S., Moshtaghie, M., Ranaie, M., & Mahmoudzadeh Tussi, P. (2023). Soundscape analysis using eco-acoustic indices for the birds biodiversity assessment in urban parks (case study: Isfahan City, Iran). Environmental Monitoring and Assessment, 195(629). https://doi.org/10.1007/s10661-023-11237-2

Lausch, A., Schmidt, A., & Tischendorf, L. (2015). Data mining and linked open data–New perspectives for data analysis in environmental research. *Ecological Modelling*, *295*, 5-17.

Lee, L. S. H., Zhang, H., Ng, K. T. K., Lo, S. C., & Yu, A. S. L. (2022). Analysing urban trees on verges and slopes along a highway using machine learning methods. Urban Forestry & Urban Greening, 78, 127786. https://doi.org/10.1016/j.ufug.2022.127786

Lotfi, R., Vaseei, M., Ali, S. S., Davoodi, S. M. R., Bazregar, M., & Sadeghi, S. (2024). Budget allocation problem for projects with considering risks, robustness, resiliency, and sustainability requirements. *Results in Engineering*, *24*, 102828.

Martins, G. B., Cué La Rosa, L. E., Nigri Happ, P., Coelho Filho, L. C. T., Santos, C. J. F., Feitosa, R. Q., & Ferreira, M. P. (2021). Deep learning-based tree species mapping in a highly diverse tropical urban setting. Urban Forestry & Urban Greening, 64, 127241. https://doi.org/10.1016/j.ufug.2021.127241

Montas, E. B. (2024). ECO-LENS: Addressing urban biodiversity with machine learning (Master's thesis, Massachusetts Institute of Technology). Massachusetts Institute of Technology. https://dspace.mit.edu/handle/1721.1/123456

Müller, N., Ignatieva, M., Nilon, C. H., Werner, P., & Zipperer, W. C. (2013). Patterns and trends in urban biodiversity and landscape design. *Urbanization, biodiversity and ecosystem services: challenges and opportunities: a global assessment*, 123-174.

Niu, C., Gutierrez, G., Cossette, L., Sadeghi, S., Portugal, M., Zeng, S., & Zhang, P. (2024). Introducing Bibliometric Analysis: A Methodological Tutorial Using Adult Online Learning Motivation Literature (2000-2022). *American Journal of Distance Education*, 1-32.

Patil, V., Patil, J., Kadam, A., Patil, A. R., Mokashi, D., & Lonare, G. M. (2024). AI-driven green space optimization for sustainable urban parks: Enhancing biodiversity and resource efficiency. Library Progress International, 44(3), 3412-3417. Retrieved from http://www.bpasjournals.com

Prodanovic, V., Bach, P. M., & Stojkovic, M. (2024). Urban nature-based solutions planning for biodiversity outcomes: human, ecological, and artificial intelligence perspectives. Urban Ecosystems, 1-12.

Rega-Brodsky, C. C., Aronson, M. F., Piana, M. R., Carpenter, E. S., Hahs, A. K., Herrera-Montes, A., ... & Nilon, C. H. (2022). Urban biodiversity: State of the science and future directions. *Urban Ecosystems*, *25*(4), 1083-1096.

Sadeghi, S. (2024). Employee Well-being in the Age of AI: Perceptions, Concerns, Behaviors, and Outcomes. *arXiv preprint arXiv:2412.04796*.

Sadeghi, S., & Niu, C. (2024). Augmenting Human Decision-Making in K-12 Education: The Role of Artificial Intelligence in Assisting the Recruitment and Retention of Teachers of Color for Enhanced Diversity and Inclusivity. *Leadership and Policy in Schools*, 1-21.

Schwarz, N., Moretti, M., Bugalho, M. N., Davies, Z. G., Haase, D., Hack, J.,…Knapp, S. (2017). Understanding biodiversity-ecosystem service relationships in urban areas: A comprehensive literature review. *Ecosystem services*, *27*, 161-171.

Seto, K. C., Güneralp, B., & Hutyra, L. R. (2012). Global forecasts of urban expansion to 2030 and direct impacts on biodiversity and carbon pools. *Proceedings of the National Academy of Sciences*, *109*(40), 16083-16088.

Silvestro, D., Goria, S., Sterner, T., et al. (2022). Improving biodiversity protection through artificial intelligence. Nature Sustainability, 5(5), 415–424. https://doi.org/10.1038/s41893-022-00851-6

Spotswood, E. N., Beller, E. E., Grossinger, R., Grenier, J. L., Heller, N. E., & Aronson, M. F. (2021). The biological deserts fallacy: cities in their landscapes contribute more than we think to regional biodiversity. *Bioscience*, *71*(2), 148-160.

Taylor, L., & Hochuli, D. F. (2015). Creating better cities: how biodiversity and ecosystem functioning enhance urban residents' wellbeing. *Urban ecosystems*, *18*, 747-762.

Tzoulas, K., Korpela, K., Venn, S., Yli-Pelkonen, V., Kaźmierczak, A., Niemela, J., & James, P. (2007). Promoting ecosystem and human health in urban areas using Green Infrastructure: A literature review. *Landscape and urban planning*, *81*(3), 167-178.

Wang, Y., Ho, I. W.-H., Chen, Y., Wang, Y., & Lin, Y. (2022). Real-time water quality monitoring and estimation in AIoT for freshwater biodiversity conservation. IEEE Internet of Things Journal, 9(16), 14366–14374. https://doi.org/10.1109/JIOT.2021.3078166

Weiskopf, S. R., Lerman, S. B., Isbell, F., & Lyn Morelli, T. (2024). Biodiversity promotes urban ecosystem functioning. *Ecography*, *2024*(9), e07366.

Wellmann, T., Lausch, A., Scheuer, S., & Haase, D. (2020). Earth observation based indication for avian species distribution models using the spectral trait concept and machine learning in an urban setting. Ecological Indicators, 111, 106029. https://doi.org/10.1016/j.ecolind.2019.106029

Willis, K. J., & Bhagwat, S. A. (2020). Biodiversity monitoring at the landscape scale: Using satellite remote sensing technologies with ground-based observations. Biological Conservation, 241, 108288.

Zhai, Z., Liu, S., Li, Z., Ma, R., Ge, X., Feng, H., Shi, Y., & Gu, C. (2024). The spatiotemporal distribution patterns and impact factors of bird species richness: A case study of urban built-up areas in Beijing, China. Ecological Indicators, 169, 112847. https://doi.org/10.1016/j.ecolind.2024.112847

Zhao, J., Zhou, Q., Li, Y., & Feng, L. (2023). Artificial intelligence applications in biodiversity monitoring: Advances and challenges. Ecological Informatics, 75, 101816.

Zhang, C., Zhan, H., Hao, Z., & Gao, X. (2023). Classification of Complicated Urban Forest Acoustic Scenes with Deep Learning Models. Forests, 14(2), 206. https://doi.org/10.3390/f14020206

Zheng, Y., Wang, Y., Wang, X., Wen, Y., & Guo, S. (2024). Managing Landscape Urbanization and Assessing Biodiversity of Wildlife Habitats: A Study of Bobcats in San Jose, California. Land, 13(2), 152. https://doi.org/10.3390/land13020152

Ziliaskopoulos, K., & Laspidou, C. (2023). Using remote-sensing and citizen-science data to assess urban biodiversity for sustainable cityscapes. Research Square. https://doi.org/10.21203/rs.3.rs-2973172/v1